# Step bunching-induced vertical lattice mismatch and crystallographic tilt in vicinal BiFeO$_3$(001) films

T. H. Kim,[1] S. H. Baek,[2] S. Y. Jang,[1] S. M. Yang,[1] S. H. Chang,[1] T. K. Song,[3] J.-G. Yoon,[4] C. B. Eom,[2] J.-S. Chung,[5,a] and T. W. Noh[1]

[1]*ReCFI, Department of Physics and Astronomy, Seoul National University, Seoul 151-747, Republic of Korea*

[2]*Department of Materials Science and Engineering, University of Wisconsin, Madison, WI 53706, USA*

[3]*School of Nano and Advanced Materials Engineering, Changwon National University, Changwon, Gyengnam 641-773, Republic of Korea*

[4]*Department of Physics, University of Suwon, Hwaseong, Gyeonggi-do 445-743, Republic of Korea*

[5]*Department of Physics, Soongsil University, Seoul 156-743, Republic of Korea*

---

a) Electronic mail: chungj@ssu.ac.kr

Epitaxial (001) BiFeO$_3$ thin films grown on vicinal SrTiO$_3$ substrates are under large anisotropic stress from the substrates. The variations of the crystallographic tilt angle and the *c* lattice constant, caused by the lattice mismatch, along the film thickness were analyzed quantitatively using the X-ray diffraction technique. By generalizing the Nagai model, we estimated how step bunching resulted in the vertical lattice mismatch between adjacent BiFeO$_3$ layers, which induced the strain relaxation and crystallographic tilt. The step bunching was confirmed by the increased terrace width on the BiFeO$_3$ surface.

Recently, there has been a great deal of interest in BiFeO$_3$ (BFO) because of its room-temperature multiferroic properties and potential device applications.[1-5] BFO is a rhombohedrally distorted perovskite, the symmetry of which determines the directions of spontaneous polarization and magnetization.[1,5] Due to its low symmetry, BFO films on a cubic substrate, such as SrTiO$_3$ (STO), should be under substantial stress associated with the epitaxial mismatch and several structural variants have been reported.[2,3] Via couplings between the lattice and other order parameters, the structural variants can affect many physical properties, including the ferroelectric remnant polarization, coercive field, and leakage current.[2]

In the growth of BFO films, vicinal substrates are commonly used. It has been reported that anisotropic strain in vicinal substrates can be used to simplify the ferroelectric domain structures in BFO(001) films.[2,3] Jang *et al*. stated that strain relaxation by preferential dislocation nucleation was responsible for the change of crystallographic tilt in vicinal BFO(001) films.[2] Using strain gradients in vicinal BFO(001) films, Kim *et al*. showed that the directional motion of the ferroelectric domain could be controlled by the polarity of the external electric bias.[4] Recently, using a synchrotron X-ray microdiffraction technique, Sichel *et al*. reported that structural relaxation resulted in BFO mosaic blocks oriented in slightly different angles.[6] However, the detailed mechanisms of the structural relaxation remain unclear and no quantitative analysis of crystallographic tilt angle $\Delta\alpha$ has been done for thicker films yet, where $\Delta\alpha$ is defined as the angle between the normals of the (001) atomic planes of STO and BFO.

In this letter, we have shown that the structural relaxation occurred through the step bunching process and lattice dislocations in BFO(001) films grown on vicinal STO substrates. Using the X-ray diffraction technique, we measured $\Delta\alpha$ and $c_{BFO}$, the $c$ lattice constant of BFO(001) films with various thicknesses. Generalizing the Nagai model[7], we could estimate the bunching rates from $\Delta\alpha$ and $c_{BFO}$. The surface morphology and the terrace width were obtained from atomic force microscopy (AFM) to confirm step bunching. As the films get thicker, step bunching becomes more likely to occur, resulting in the increase of the terrace width and $\Delta\alpha$. The relationship between the bunching rate and the terrace width quantitatively agrees with the predictions of the step-bunching instability model of Tersoff *et al.*[8]

High-quality Pt (40 nm)/BFO (50, 100, 200, and 400 nm)/SrRuO$_3$ (SRO) (50 nm) heterostructures were grown epitaxially on vicinal STO(001) substrates with a miscut angle of $\alpha \approx 2$ and 4° along the [100] direction.[2,9] The structural properties of vicinal BFO(001) films were investigated using high-resolution X-ray diffractometer. For each film, the miscut direction was initially determined from the orientation change of the STO(002) diffraction peak with the rotation of the specimen around the surface normal (see Supplementary Fig. S1). As shown in Fig. 1, we chose [100] as the downhill miscut direction and [001] as the *c*-axis of the SRO layer, which is the same as that of the STO substrate. Then, we performed reciprocal space mapping (RSM) around the STO {002} Bragg peaks. We used two configurations, *i.e.* the phi $\phi$-angles of 0 and 90°, where the [100] and [010] directions lie in the X-ray incidence plane, respectively. From these RSMs, we determined $\Delta\alpha$ and $c_{BFO}$. Independently, the averaged terrace width and the step height on the top surfaces were also measured using AFM to confirm the step bunching.

Figure 1 shows a schematic diagram of how the crystallographic tilt of the BFO layer evolved during film growth due to step bunching and lattice dislocation. In the initial stage, the normal of the BFO layer (the black arrow) was tilted negatively against those of the STO substrate (the dashed arrow) and the film surface (the dotted arrow). It should be noted that $\alpha$ was defined as the angle between the normals of the film surface and the STO(001) plane. The vicinality of the underlying substrate generated strong in-plane and out-of-plane compressive strains, which tilt the BFO layer to a negative $\Delta\alpha$ with respect to the STO substrate (the BFO region with the red hatched patterns near the substrate). The negative crystallographic tilt of the BFO layer was explained well by the original Nagai model, which assumed 1:1 matching between a layer and the sublayer at the steps:[7]

$$\tan \Delta \alpha = -\frac{c_{\text{layer}} - c_{\text{sublayer}}}{c_{\text{sublayer}}} \cdot \tan \alpha_{\text{sublayer}} = -\frac{c_{\text{layer}} - c_{\text{sublayer}}}{c_{\text{substrate}}} \cdot \tan \alpha \qquad (1)$$

where $c_{\text{layer}}$, $c_{\text{sublayer}}$, and $c_{\text{substrate}}$ are the $c$ lattice constants of a layer, the less relaxed sublayer, and the underlying substrate, respectively. For the first BFO layer, $c_{\text{sublayer}}$ and $c_{\text{substrate}}$ are the $c$ lattice constants of the SRO layer and the STO substrate, respectively. Note that we used $c_{\text{substrate}}$ as the denominator of Eq. (1) by the definition of $\alpha = \alpha_{\text{STO}}$.

As the BFO film grows, imperfect matching between adjacent BFO layers at step edges can occur due to step bunching, *i.e.* coalescence of isolated steps induced by thickness-dependent strain relaxation.[8] As shown in Fig. 1, on the step edges, the lattice mismatch forces a few BFO unit cells on the right side (marked with blue checked patterns) to have locally larger $c$ lattice parameters than the neighboring unit cells on the left side (marked with red hatched lines), where they were relaxed close to the value of bulk BFO by introduction of in-plane misfit dislocations. The in-plane misfit dislocations do not cause tilt of (001) planes. On the other hand, the lattice mismatch along [001] at the step edge generates vertical misfit dislocations, which results in tilt of atomic planes. Eventually, dislocation-mediated structural relaxation results in mosaic blocks in slightly different orientations, as reported by Sichel *et al.*[6] But, on average, the gradual change of the [001] strain on terraces makes the normal of the BFO layer in blue checked region rotate toward that of the film surface, resulting in a positive crystallographic tilt.

From {002} RSMs of vicinal BFO(001) films, we found that $\Delta\alpha$ changed from a negative to a positive value as the film thickness increased. For a 50-nm-thick BFO film, the RSM in Fig. 2(b) showed that all of the STO, SRO, and BFO peaks were located along the vertical line, indicating no tilt in the [010] direction. On the other hand, the RSM in Fig. 2(a) showed that the BFO peak was located on the right side of the vertical line, indicating negative tilt in the [100] direction. From the angle between the vertical line and BFO peak, $\Delta\alpha$ was estimated as -0.15°, in good agreement with the value of -0.12° calculated using Eq. (1). Thus, the negative crystallographic tilt was explained well by the Nagai model. However, for a 400-nm-thick film, as shown in Figs. 2(c) and 2(d), $\Delta\alpha$ was a positive value of 0.64°.

By assuming the vertical matching ratio $r_{match}$ between adjacent film layers at the terrace edge, it was possible to extend the simple Nagai model as:[10]

$$\tan\Delta\alpha = \frac{r_{match} \times c_{sublayer} - c_{layer}}{r_{match} \times c_{sublayer}} \cdot \tan\alpha \qquad (2)$$

Here, $r_{match}$ is the vertical matching ratio. When step bunching occurs, the $m$ strained film layers will be vertically matched with the ($m$ + 1) relatively relaxed film sublayer at the bunched step edges and $r_{match}$ is represented as $(m + 1)/m$. When $r_{match} = 1$, this equation becomes the original Nagai model (*i.e.*, Eq. (1)). When $r_{match} > 1$ in mismatched epitaxial films, Eq. (2) predicts the positive crystallographic tilt.

By applying the extended Nagai model, we calculated the corresponding $r_{match}$ and $m$ (*i.e.*, the bunching rate) from $\Delta\alpha$ and $c_{BFO}$ in BFO(001) films with different thickness (Table I). For $c_{sublayer}$, $c_{BFO}$ of a thinner film was used. For example, $c_{BFO}$ of the 200-nm-thick film was used as $c_{sublayer}$ of the 400-nm-thick film. As the film got thicker, $c_{BFO}$ and $m$ decreased, suggesting that the compressive strain from underlying substrates became relaxed by the step bunching process. Note that $m$ should be larger than the film thickness for a fully strained film, as the case for the 50-nm-thick BFO film, which was about 125-unit-cell-thick. The value $m$ is an integer only when the $r_{match}$ value does not change throughout the film. However, in reality, $r_{match}$ may change as the film grew. Then, the $m$ value can be a non-integer and represents an average.

The step bunching process decreases the number of steps and consequently, increases the distance between steps. We independently measured the terrace widths and the step heights of the BFO top layers from AFM images (see Supplementary Fig. S2). As shown in Fig. 3, the terrace width increased as the film got thicker, confirming the step bunching.

Tersoff *et al.*[8] developed a theory for the step-bunching instability of a vicinal surface under stress. They showed that the surface under stress is always unstable against step bunching, due to a long-range attraction between the steps induced by elastic relaxation, causing the progressive coalescence of steps. As the distance between steps increases, the bunching rate decreases proportionally nearly to the inverse third power of the average step separation. The predicted line from the model by Tersoff *et al.*[8] is shown as the dotted curve in Fig. 3. Here, we assumed the minimum-energy separation for step bunching as the terrace width of the 50-nm-thick BFO(001) film, which was free from the step bunching (*i.e.*, $r_{match}$ ~ 1). Since the minimum-energy separation depends only on bulk properties, our experimental data agree with the theoretical predictions excellently regardless of the $\alpha$ values, suggesting the importance of the step bunching process in the strain relaxation of vicinal BFO(001) films.

In summary, we showed that positive crystallographic tilt of a vicinal BiFeO$_3$(001) film surface was caused by step bunching processes. By generalizing the Nagai model, we related how the vertical lattice mismatch between bunched BiFeO$_3$ layers resulted in a positive crystallographic tilt. The step bunching was confirmed by the increased terrace width and the step height on the top surface. Our work increased our understanding of the physical properties of the epitaxial films on the vicinal substrates and can be used when growing various epitaxial films on the vicinal substrates.

This work was supported by National Research Foundation of Korea (NRF) grants from the Korean government (MEST) (Grants 2009-0077249: J.-S.C., 2009-0080567 and 2010-0020416: T.W.N.). The work at University of Wisconsin-Madison was supported by the Army Research Office under Grant No. W911NF-10-1-0362 (C.B.E.).

TABLE I. Changes in $\Delta\alpha$, $c_{BFO}$, $m$, and the terrace width at the top surface with respect to the thickness of vicinal BFO(001) films with different $\alpha$ values.

| Thickness (nm) | $\alpha, \Delta\alpha$ (°) | $c_{BFO}$ (Å) | $m$ | Terrace width (nm) |
|---|---|---|---|---|
| 50 | 4, -0.15 | 4.067 | 161 | 204 |
| 100 | 4, 0.25 | 4.031 | 14.3 | 256 |
| 200 | 4, 0.38 | 4.014 | 10.0 | 285 |
| 400 | 4, 0.64 | 3.990 | 5.26 | 313 |
| 200 | 2, 0.72 | 4.004 | 1.96 | 450 |

FIG. 1. (Color online) Schematic diagram of step bunching-induced vertical lattice mismatches in vicinal BFO(001) films. Distorted, open, and shaded blocks represent the BFO layer, the SRO bottom electrode, and the STO substrate, respectively. Black, dashed, and dotted arrows represent the normals of the BFO layer, the STO substrate, and the film surface, respectively. The monoclinic distortion of BFO unit cells was toward the downhill miscut direction of [100]. The BFO blocks with red hatched (blue checked) patterns represent regions where the normal of the BFO layer is negatively (positively) tilted relative to that of the STO substrate. At a bunched step edge, three blue checked BFO layers were vertically matched with four red hatched BFO sublayers.

FIG. 2. (Color online) RSMs at $\phi = 0$ and 90° around STO {002} Bragg peaks in (a, b) 50- and (c, d) 400-nm-thick BFO films grown on vicinal STO(001) substrates with a 4° miscut angle. The directions of red arrows in (a) and (c) indicated that the BFO layers were crystallographically tilted with negative and positive angles, respectively.

FIG. 3. (Color online) The values of $m$ vs. the normalized terrace widths in vicinal BFO(001) films with various thicknesses. The dotted curve shows the correlation between $m$ and the normalized terrace width. The terrace widths on the top surface of the 100-, 200-, and 400-nm-thick BFO(001) films were normalized by that of a 50-nm-thick film. The solid blue star shows that our picture also holds for 200-nm-thick BFO film grown on a STO(001) substrate with a 2° miscut angle.

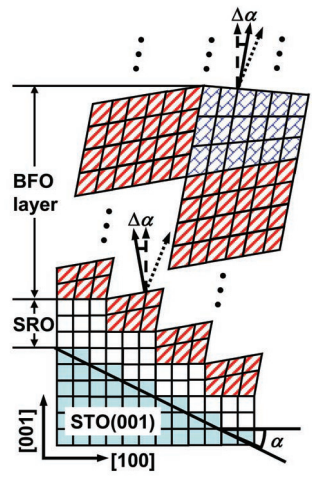

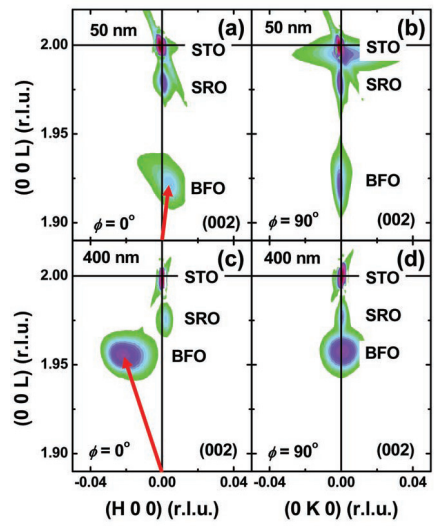

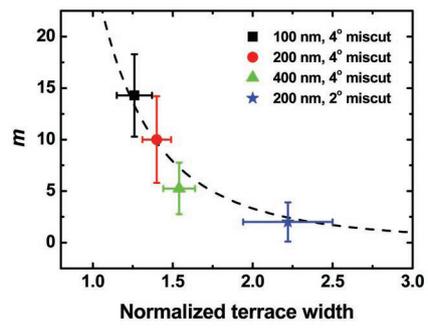